\documentclass[aps,epsf,preprint]{revtex4}
\usepackage{amssymb}
\usepackage{color}

\def\1{{\bf 1}}
\def\[{\left[}
\def\]{\right]}
\def\be{\begin{eqnarray}}
\def\ee{\end{eqnarray}}
\def\bm{\begin{pmatrix}}
\def\em{\end{pmatrix}}
\def\nn{\nonumber}
\def\({\left(}
\def\){\right)}
\def\bk#1{\langle#1\rangle}
\def\eq#1{(\ref{#1})}

\def\a{\alpha}

\def\s{\sigma}

\def\p{\partial}

\def\labels#1{\label{#1}}
\def\edc{\end{document}}

\def\bn{\begin{enumerate}}
\def\i{\item}
\def\en{\end{enumerate}}
\def\b{\beta}

\def\ba{\begin{array}}
\def\ea{\end{array}}
\def\bc{\begin{center}}
\def\ec{\end{center}}

\def\edoc{\end{document}}

\def\^{$\wedge$}
\def\.{\!\cdot\!}

\def\igw#1{\includegraphics[width=#1cm]}
\def\+{\!+\!}
\def\-{\!-\!}

\def\bsl{\backslash}

\def\sb#1{\!\stackrel{.}{#1}\!}

\def\sa#1{\stackrel{#1}{\longrightarrow}}

\def\sp#1{{1\over k_{#1}^2}}
\def\1{\sb 1}
\def\2{\sb 2}
\def\3{\sb 3}
\def\4{\sb 4}
\def\5{\sb 5}
\def\6{\sb 6}
\def\7{\sb 7}
\def\8{\sb 8}
\def\9{\sb 9}
\def\M{M\"obius\ }
\usepackage{graphicx}
\graphicspath{/Users/harrylam/Current/filetex/Helicity/graphics/}
\def\S{Sec.~}
\begin{document}
\title{The Role of M\"obius Constants and Scattering Functions in CHY Scalar Amplitudes}
\author{C.S. Lam$^a$  and York-Peng Yao$^b$}
\address{$^a$Department of Physics, McGill University\\
 Montreal, Q.C., Canada H3A 2T8\\
$^a$Department of Physics and Astronomy, University of British Columbia,  Vancouver, BC, Canada V6T 1Z1 \\
$^b$Department of Physics, The University of Michigan
Ann Arbor, MI 48109, USA\\
Emails: Lam@physics.mcgill.ca\  yyao@umich.edu}

\begin{abstract}
The integration over the M\"obius variables
 leading to the Cachazo-He-Yuan (CHY) double-color $n$-point massless scalar amplitude are
carried out one integral at a time. \M invariance dictates the final amplitude to be independent of the three \M
constants $\s_r, \s_s, \s_t$, but their choice affects integrations and the intermediate results. 
The effect of the \M constants, which will be held finite but otherwise arbitrary, 
the two sets of colors, and the scattering functions on each integration is investigated.
A general systematic way to carry out the $n-3$ integrations is explained, each exposing one of the $n-3$ propagators
of a single Feynman diagram. Two detailed examples are shown to illustrate the procedure, one a five-point amplitude,
and the other a nine-point amplitude. Our procedure does not generate intermediate spurious poles, in contrast to what is common by choosing Mšbius constants at 0, 1, and $\infty$. 
\end{abstract}
\narrowtext
\maketitle

\section{Introduction}
The recently discovered scattering formula of Cachazo, He, and Yuan (CHY) \cite{CHY1,CHY2,CHY3,CHY4, CHY5, CHY6}
opens up a new chapter on  the formulation of particle scattering amplitudes. 
It resembles string theory in its formulation \cite{ST1,ST2,ST3}, but it works with any space-time dimension, including four,
without supersymmetry, nor any string-scale excitation. It reproduces ordinary Feynman diagrams in the tree
approximation for massless scalar, gauge, and graviton scatterings, without a Lagrangian nor  any quantized local field 
operator. Its only connection with space-time is through the external momenta, 
and in the case of spin-1 and spin-2 particles, also through their polarization vectors. These
space-time vectors enter as dot products so Lorentz invariance is automatically guaranteed. In this formulation,
dynamics is introduced not by a local interaction Lagrangian with  Lorentz invariance, but by
insisting on a M\"obius invariance of the scattering amplitude in an underlying
complex space.
The fact that it reproduces Feynman diagrams obtained from quantum field theory,
despite the fact that neither local interaction nor quantization rules is explicitly put in,
suggests that both quantum uncertainty and local interaction are to some extent emerging, not
introduced by hand. This promise of a novel approach to a well-known theory is certainly worth more detailed
scrutiny and further investigation.

It is true that there is still much to be learned. In order to implement unitarity we need to know how
loops are expressed in the CHY formalism \cite{LOOP1,LOOP2,LOOP3,LOOP4}, 
or how the scattering theory is related to quantum field theory
where hermiticity of its Hamiltonian formally guarantees unitarity. As a first step 
one needs to find out how to express an off-shell scattering amplitude or a Green's function in the CHY 
formalism, a task which is carried out in Ref.~\cite{LY1}. 
One also needs to know how to apply the formalism to other theories \cite{MASS1,MASS2,MASS3,MASS4},
including the Standard Model. 

In the present article we study the simplest of these theories, the massless double-color scalar amplitude.
This amplitude can be expressed as a sum over the $(n\-3)!$ solutions of the scattering equations
\cite{CHY3}, but since these solutions are roots of polynomials of degree $(n\-3)!$ \cite{DG2,CK15,DG3}, whose
analytic solution is not known for $n>5$, it is difficult to compute it this way beyond $n=4$ and $n=5$
\cite{CHY3,Kal}. However, starting from these small $n$ amplitudes which have a Feynman-diagram interpretation, 
it is possible to generalize the Feynman-diagram correspondence to all $n$, using trivalent graphs \cite{CHY3,FG}, or
 polygon graphs \cite{BBBD2}, or the pairing diagrams to be explained in Sec.~IV. The advantage of the pairing diagram
 is that it deals directly with Feynman diagrams, without going through an intermediate trivalent or polygon graph.

Alternatively, the amplitude can be written as an $(n\-3)$-dimensional complex integral.
With the help of the global residue theorem for two or more complex variables \cite{DG1,BBBD}, 
$n=5$ amplitude can be computed, and the
Britto-Cachazo-Feng-Witten (BCFW) factorization \cite{BCFW} can be established.

The trivalent and the polygon graphs, or the pairing diagrams, are determined
solely from the two colors $\a$ and $\b$, without the involvement of scattering functions $f_i$. Nevertheless,
scattering functions must be crucial to the amplitude
because they are the only source of momentum in the CHY formula. To investigate
their role and how unique they must be,
we need to carry out  the $(n\-3)$ integrations  in the CHY amplitude,  one at a time. 
We shall find that a mechanism discussed in \S III morphs the scattering functions into Feynman propagators.
We shall also find  that the scattering functions are flexible enough to provide correct
momentum dependence in every integration.

The theory of function of several complex variables is much more complicated than the
theory for one complex variable. For example, a function of one variable analytic inside a smooth region can 
always be represented by a Cauchy integral on its boundary, but this is no longer true for a function of 
several complex variables. Most of the time, a function analytic in a region is automatically analytic in a much
larger region, the natural domain of holomorphy, and the generalization of a Cauchy integral is the Bergman-Weil
integral taken over the boundary of a natural domain of holomorphy. This is why 
the  K\"all\'en-Lehmann representation of a two-point function is much simpler than the corresponding
K\"all\'en-Toll representation of a three-point function \cite{KT}. Although things are much simpler for meromorphic
functions, even so it is still non-trivial to calculate the value and the sign of multi-dimensional residues.
For that reason it is much safer to perform the $(n\-3)$ integrations one at a time, 
a tactic which we will follow in the rest of this article.

In carrying out each of these integrations, the location of the constant lines $r,s,t$ become relevant,
so their role has to be understood. 
As a result of \M invariance, the final amplitude in \eq{m} is independent of the values of three \M constants, 
$\s_r, \s_s, \s_t$, nor the choice of the three constant lines $r,s,t$, but the complexity and the result of  intermediate
integrations do depend on their choice. 
 In this respect the choice of \M constants
is like the choice of a gauge in a gauge theory. 
The complexity
of the intermediate computation depends on  the gauge, but the final scattering amplitude does not.
It would have been much simpler if we did not have to choose a gauge, but unfortunately this cannot be done in concrete calculations.

Although the formulations are different, technically there is a certain amount of overlap with Refs.~\cite{DG1,BBBD,BBBD2}, but 
there are also notable differences.
They chose a gauge in which $\s_{r,s,t}=0,1,\infty$, but we leave their values arbitrary. There are two advantages in doing so. The necessary
cancellation of $\s_{r,s,t}$ at the end gives  a way to check the correctness of the algebraic manipulation in between. Moreover, in the
gauge $\s_{r,s,t}=0,1,\infty$, the leg with an infinite \M constant disappear from the integrand, creating an asymmetry. In the case of $n=5$ 
with diagonal color $\a=\b$, for example, if we compute it in a general gauge, there are five dominant regions of integration giving rise directly to
the five Feynman diagrams, whereas in the $\s_{r,s,t}=0,1,\infty$ gauge, there are three dominant regions of integrations, giving rise fictitious poles,
which get cancelled only after all the terms are combined. See Appendix A for details.

We also discuss in detail how the dominant regions of integration (called a `multi-crystal' later) are affected by the choice of $r,s,t$. 
Each dominant region of integration consists of non-overlapping sub-regions, and/or sub-regions with one completely inside anther. 
Each non-overlapping
sub-region must contain one of $r,s,t$, but depending on what they are, one or two of these three lines may never be able to
 enter any dominant integration
region at all. Some illustrative examples for this intricacy can be found in \S XI.

In a separate publication, we will apply what is learned about doing integration here to the evaluation of the CHY gauge amplitude \cite{LYgauge}.

How individual integrations should be carried out is explained in \S II. 
After introducing some useful terminologies
and results in Secs.~III, IV, V, and VI, how integrations can be carried out is explained in \S VII.
As a result, the amplitude \eq{m} is factorized into two `partial amplitudes', joined by a propagator which
originated from an inverse scattering function. The partial amplitudes resemble the CHY amplitudes but are
not the same, and the two partial amplitudes are also somewhat asymmetrical. See \S XI for more
discussions. Nevertheless, we shall show
in Sec.~IX that the resemblance is sufficiently close to allow the partial amplitudes to be factorized again
and again in subsequent integrations, 
until all the propagators are exposed. A nine-point CHY amplitude is worked out in detail in
\S X to illustrate the procedure, and a summary section is included in \S XII. 

To explain these details, many mathematical notations are required which unfortunately
seem to make the discussion somewhat complicated. 
However, using the notions introduced in Secs. IV and V, an intuitive physical picture of
crystal fracture can be
concocted to describe these mathematical operations. This is explained in Sec. VIII.

\section{Double Colored Scalar Amplitude}
The color-stripped massless scalar tree amplitude is given by the CHY formula \cite{CHY3} to be
\be m(\a|\b)=\({-1\over 2\pi i}\)^{n-3}\oint_\Gamma \s_{(rst)}^2\(\prod_{i\in A\bsl r,s,t}{d\s_i\over f_i}\)
{1\over\s_{(\a)}\s_{(\b)}},\labels{m}\ee
where $\a=(\a_1\a_2\cdots\a_n)\in S_n$ and $ \b=(\b_1\b_2\cdots\b_n)\in S_n$ are two {\it cyclically ordered} sets
describing  colors,  $A=\{12\cdots n\}$ is the  set of external lines, and
$\Gamma$ is a contour  in the $(n\-3)$-dimensional complex space 
encircling the $(n\-3)$ $f_i=0\ (i\!\not=\!r,s,t)$.
For every $i\!\in\! A$, the scattering function $f_i$ is defined to be
\be
f_i&=&\sum_{j\in A\backslash i}^n{2k_i\.k_j\over\s_{ij}},\quad \sum_{i\in A}k_i=0,\labels{sf}\ee
and the $\s$'s are given by
\be
\s_{ij}&=&\s_i-\s_j,\quad\s_{(\a)}=\s_{(\a_1\a_2\cdots\a_n)},\quad \s_{(\b)}=\s_{(\b_1\b_2\cdots\b_n)},\nn\\
\s_{(abc\cdots de)}&=&\s_{ab}\s_{bc}\cdots\s_{de}\s_{ea}\equiv \s_{[abc\cdots de]}\s_{ea}.\labels{sigma}\ee
Note that $\s_{ij}=\s_{[ij]}$ differs from $\s_{(ij)}$ by a factor $\s_{ji}$.

The factor $\s_{(\a)}\s_{(\b)}$ governs dynamics and
 is replaced by something else in  gauge scattering and  in gravitational scattering.
For that reason we will refer to it as the {\it dynamical factor}.

It is known that if all the external momenta $k_i$ are massless (on-shell), then every $f_i$ transforms covariantly
under a M\"obius transformation $\s_j\to (a\s_j+b)/(c\s_j+d)$,  $ad-bc=1$. In that case
$f_i\to f_i(c\s_i+d)^2$, and  the amplitude $m(\a|\b)$ is M\"obius invariant. Consequently \eq{m} is independent
of  the M\"obius constants $\s_r, \s_s$, and $\s_t$ and the choice of the constant lines $r,s,t$. Moreover, 
the functions $f_r, f_s, f_t$ are automatically  zero  when the  variable scattering
functions $f_i\ (i\not=r,s,t)$  vanish.

A straight-forward evaluation of \eq{m} at the poles enclosed by $\Gamma$
requires  explicit solutions of  the scattering equations $f_i=0$, which are hard to obtain for $n>5$.
A simpler method is to  distort the contour away from some specific $f_p=0$ 
and to evaluate the integral at the poles defined by
$\s_{(\a)}\s_{(\b)}=0$, but it is safe to do so only one complex variable at a time because the 
theory of functions of several complex variables is rather complicated. 
This forces us to choose an ordering of these individual integrations. Whatever it is, each integration exposes one of the
$(n\-3)$ propagators of a Feynman tree diagram.
We will provide an evaluation of the integral along this line, in the spirit of  \cite{DG1}, but different in details.
The BCFW-factorization technique is not used here,  neither is the total residue theorem for two and more variables.
The M\"obius constants $\s_r, \s_s, \s_t$ are left arbitrary, not set 
equal to $0, 1, \infty$. 
The final result is the same as obtained previously, but the role of M\"obius constants and order of integration are made 
clear  this way. Moreover, the cancellation of  $\s_r, \s_s, \s_t$ dependences at the end often require algebraic combinations and
manipulations. Algebraic mistakes could lead to non-cancellation, so keeping them around and arbitrary is a 
good way to check the algebra.

This approach also exposes 
a parallel between quantum field theory and the CHY formula. In quantum field theory, the
d'Alambertian operator $\p^2$ in Klein-Gordon equation turns into an inverse propagator when it goes off-shell. In the CHY
theory, the scattering function $f_p$ also turns into an inverse propagator when
it moves away from the scattering equation $f_p=0$.

Before embarking on such an evaluation, let us  first explain some terminologies which will be useful later.

\section{Partial scattering functions}
It would be useful for later purpose to define {\it partial scattering functions}  for a subset $S\subset A$ to be
\be
f_a^S=\sum_{b\in S\bsl a}{2k_a\.k_b\over\s_{ab}},\quad (a\in S).\labels{psf}\ee
The total momentum of this subset is generally not zero. Its deficit is denoted by
$k^S=-\sum_{b\in S}k_b$. In particular, $k^A=0$ and $f^A_i=f_i$.

The following sum rule holds for the partial scattering functions, whatever $\s_a$'s are,
and whether $k_a$ are on-shell or not.
\be
\sum_{a\in S}f_a^S&=&0,\nn\\
\sum_{a\in S}\s_af_a^S&=&(k^S)^2-\sum_{a\in S}k_a^2.\labels{ps}\ee
The first identity follows from the antisymmetry of $a$ and $b$ in the equation 
$\sum_{a\in S}f_a^S=\sum_{a,b\in S, b\not=a}{2k_a\.k_b/\s_{ab}}=0$. The second identity is true because
$$\sum_{a\in S}\s_af_a^S=\sum_{a,b\in S, b\not=a}{(\s_{ab}+\s_b)2k_a\.k_b\over\s_{ab}}=-\sum_{a\in S}2k_a\.(k^S+k_a)
-\sum_{b\in S}\s_bf_b^S,$$
hence
$$\sum_{a\in S}\s_af_a^S=(k^S)^2-\sum_{a\in S}k_a^2.$$

In particular, if every particle in $S$ is on-shell and $f_a^S=0$ for every $a\in S\bsl p,r$, then $f_p$ and $f_r$ can be obtained from \eq{ps} to be
\be  f_p^S=-f_r^S={(k^S)^2\over\s_{pr}}.\labels{fpr}\ee

\section{Color Pairings}
It turns out that integrations in \eq{m} are controlled by  pairings of the two colors $\a$ and $\b$.
These pairings are determined algebraically, but may also be expressed graphically as `pairing diagrams'. 
The latter is more intuitive and more convenient because they turn directly into Feynman diagrams of the amplitude.
However, it should be emphasized that pairing diagrams are obtained purely from  colors $\a$ and $\b$,
without invoking dynamics. The reason why they finally coincide with the Feynman diagrams comes from the structure
of the CHY formula, not needed at this stage.

Two lines that are adjacent both in $\a$ and in $\b$ constitute  a {\it pair}. For example, 
if $\a=(123456789)$ and $\b=(124395786)$, then the pairs are $\bk{12}, \bk{34}$, and $\bk{78}$. We will refer to these primordial pairs
as {\it level-1} pairs. Now mentally merge the level-1 pairs into a single unit and look for new
pairs in $\a$ and $\b$ to form  {\it level-2} pairs.
In the example above,  $\bk{\bk{12}\bk{34}}$ and
$\bk{6\bk{78}}$ are level-2 pairs. In a similar manner,  {\it level-$\ell$} pairs are new pairs obtained by mentally merging 
all lower-level pairs 
into a single unit. We say that $\a$ and $\b$ are totally paired if all the lines in them are paired up this way into one of two groups.

There is no guarantee that any two color can be totally paired. 
It is also possible that two colors can be totally paired in more than one way.

In the example above, the  total pairing is $\langle 9\langle \langle 12\rangle \langle 34\rangle \rangle \rangle \langle 5\langle 6\langle 78\rangle \rangle \rangle $ in the $\a$-perspective,
and $\bk{\bk{\bk{12}\bk{43}}9}\bk{5\bk{\bk{78}6}}$ in the $\b$-perspective. Final pairing depends on
how  higher-level pairings are executed so it is not unique. A line belonging to one group may be shifted to another group.
 For example,  another possible total pairing  is $\bk{\bk{12}\bk{34}}\bk{\bk{5\bk{6\bk{78}}}9}$ in the $\a$-perspective and $\bk{\bk{12}\bk{43}}\bk{9\bk{5\bk{\bk{78}6}}}$ in the $\b$-perspective, where line 9 has been shifted from one group to another.

\bc\igw{6}{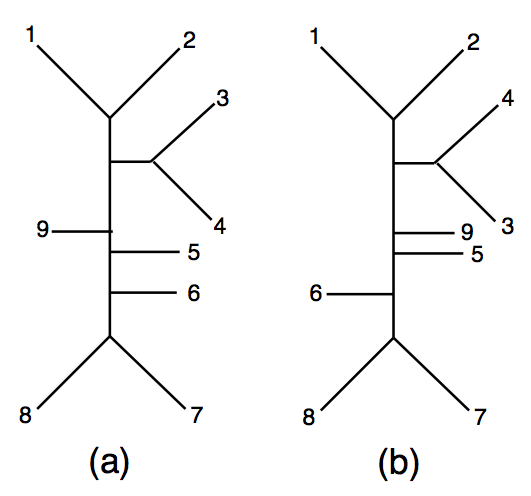}\\ Fig.~1.\quad Pairing diagrams of two colors $\a$ and $\b$. (a), 
in the $\a$-perspective; (b), in the $\b$-perspective\ec

Final pairings of $\a$ and $\b$ can be represented by  {\it pairing diagrams}, one in the $\a$-perspective and  one
in the $\b$-perspective. For example, the pairing diagrams for the example above are shown in Fig.~1.
External lines of the two diagrams, read cyclically clockwise,
give the two colors $\a$ and $\b$ respectively. 
Level-1 pairs are vertices with two external lines. 
Each bracket $\bk{\cdots}$ in a pairing maps to an internal line. For example, lines 1,2,3,4 in Fig.~1 merge
into an internal line which corresponds to the outermost bracket of $\bk{\bk{12}\bk{34}}$ or $\bk{\bk{12}\bk{43}}$.
Each internal
line connects two complementary pairs, which upon being cut will
separate the whole graph into two halves. These are the two final groups discussed before. Cutting different internal
ones corresponds to a different arrangement of external lines into two groups. 
The level of a pair is simply the number of internal lines (including the one that has been cut) traversed before
 an external (level-1) pair is reached.
 
There is no need to exhibit both pairing diagrams. If one is displayed, the other can be obtained simply by demanding that
the second color and its diagram give rise to no new pairing not already contained in the original
diagram. 

Since Fig.~1(b) must contain the same pairings as Fig.~1(a), it must be obtainable from Fig.~1(a) by
flipping some lines.
Start from Fig.~1(a), line 3 must be flipped to get Fig.~1(b) because 
otherwise both $\a$ (read from 1(a)) and $\b$ (read from 1(b)) contain $(123\cdots)$, so $\bk{23}$ would be a pair 
which is not present in Fig.~1(a). Similarly,  9 must be flipped because the pair $\bk{91}$ is not
present, 6 must be flipped because $\bk{67}$ is not present. With 6 flipped, 5 cannot be flipped because $\bk{56}$ is not a 
pair. In this way 
we obtain Fig.~1(b). 

This example gives rise to one (set of) pairing diagram, but there are color pairs which produce several pairing diagrams because
there are several ways of pairing up the lines. For example, if $\a=\b=(12345)$, then there are five pairing diagrams, shown Fig.~2,
corresponding to the pairings $A=\bk{\bk{12}3}\bk{45}$, $B=\bk{\bk{23}4}\bk{51}$, $C=\bk{\bk{34}5}\bk{12}$,
$D=\bk{\bk{45}1}\bk{23}$, $E=\bk{\bk{51}2}\bk{34}$.

\bc\igw{9}{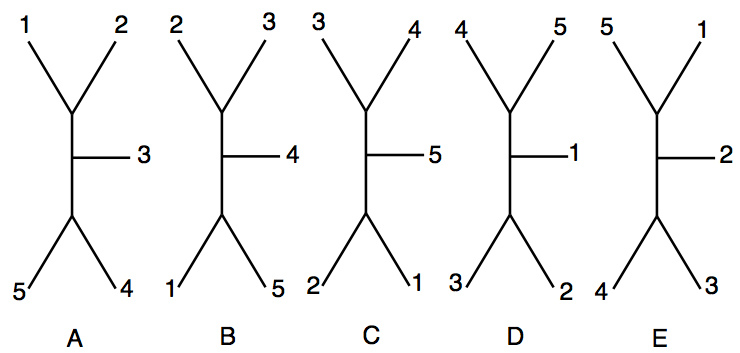}\\ Fig.~2.\quad The five pairing diagrams for $\a=\b=(12345)$\ec

\section{Final result of the CHY amplitude}
It is  known \cite{CHY3,DG1,Kal,FG,BBBD2} that  amplitude \eq{m} can be
expressed in  Feynman diagrams. These Feynman diagrams  are identical to the pairing diagrams.
If $\a$ and $\b$ can be totally paired in several ways, then each pairing gives rise to a Feynman diagram and all of them
should be summed to get the final result. If $\a$ and $\b$ cannot be totally paired, then the amplitude \eq{m} is zero.
The overall sign of the amplitude is determined by the relative signature of $\a\in S_n$ and $\b\in S_n$.

In this article we shall arrive at these results by carrying out the $(n\-3)$ integrations, one after another.

To start with, it seems surprising that pairing diagrams alone can determine the final amplitude. After all,  momentum dependence of
the amplitude can only come from the scattering functions $f_i$, but  they are not even needed in determining the pairing diagrams.
Moreover, three arbitrary constant lines $r,s,t$
and three arbitrary \M constants $\s_r, \s_s, \s_t$ enter into the formula of $m(\a|\b)$.
Although M\"obius invariance dictates the final amplitude 
to be independent of the their choice, they and $f_i$ must somehow get involved in  intermediate computations. How these quantities
affect the individual integrations,  and how the final independence on $\s_r, \s_s, \s_t$ comes about,
are the central points to be investigated in this paper.

To facilitate further 
discussion, we  introduce the terminology `crystals' and `defects', to incorporate  knowledge on the location of  constant lines
needed to carry out integrations into  pairs.

To simplify writing, we will assume without prejudice  from now on
that $\a=(123\cdots n)$. Any other $\a$ can be obtained by relabeling the lines.

\section{Crystals and Defects}
A {\it crystal} is a set of consecutive lines forming a pair at some level, provided it contains one and only one constant line.
In a pairing diagram, a crystal set is made up of the external lines to one side of an internal line, if there is also a constant line among them.
The constant line in this set is called the {\it defect} of the crystal. These names are chosen
to give an intuitive feeling of the integration process, as will be described in \S IX.

 For example, if we take 2,6,9 to be the constant lines in Fig.~1, then its crystals are $\{12\}, \{1234\}, \{678\}$,
 and $\{5678\}$, and no more. Their defects  are respectively 2, 2, 6, 6, and there is no crystal containing the the constant line 9 as a defect.
 Note that $\bk{34}$ is a pair but $\{34\}$ is not a crystal, because it does not contain a constant line. Hence the number of crystals
 is generally less than the number of pairs.

Crystals are important because they are the basic integration units. 

\section{Amplitude factorization}
In the present and the following sections, we shall show that a single integration of
the amplitude in \eq{m} results in a factorization of the amplitude into a product of two 
`partial amplitudes', connected by a propagator, as shown in Fig.~3.
Partial amplitudes are similar to but not the same as amplitudes shown in \eq{m}. Nevertheless, they  can also be factored
by further integrations,  over and over again, 
until all propagators of
the Feynman-diagram are exposed. 

\bc\igw{4}{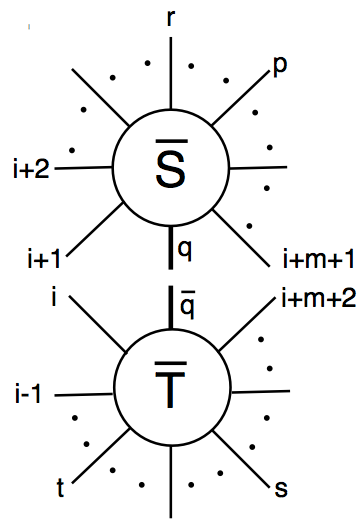}\\Fig.~3.\quad Factorization of the amplitude in equation \eq{m}\ec

To make it easier to explain and simpler to write, we shall assume
 $\a=(1234\cdots n)$. This poses no restriction because a relabeling of the lines can accommodate any other $\a$.
 
As mentioned previously, the idea  is to move the contour $\Gamma$ in \eq{m} away from 
$f_p=0$ for a single $p$, but  keep it encircling  the remaining $f_i=0\ (i\in A\bsl r,s,t,p)$.
Distort the contour until it surrounds the poles of $1/\s_{(\a)}\s_{(\b)}$, then the
result of the integration is  simply $-2\pi i$ times the residues at these poles. In this way the necessity of obtaining 
complicated solutions of the scattering equations \cite{SE1,SE2,SE3,SE4,SE5,SE6,SE7,SE8,SE9}  is avoided.

It turns out that the integrand has a pole only when the degree of zeros of $\s_{(\a)}\s_{(\b)}$ is maximal. 
For $\a=(12\cdots n)$, the degree of zeros of $\s_{(\a)}$ is $m$ when the $\s$'s of $(m\+1)$ consecutive lines
coincide. Generally the degree of zeros for $\s_{(\b)}$ at that coincidence is less than $m$, unless these $(m\+1)$
lines are paired up with $\a$ in the sense of \S~IV, in which case the degree is $m$. This happens when these $(m\+1)$
consecutive lines form a pair (at some level).

The set  $S=\{i\+1, i\+2,\cdots,i\+m, i\+m+1\}$ of $m\+1$ consecutive lines contains only
$m$ difference variables $\s_{x,x+1}\ (i\+1\le x\le i\+m)$. That suggests  that the remaining $\s$ in $S$ should
be a M\"obius constant, say $\s_r$. If this were not the case,  we shall see that a pole would not materialize.
Therefore, if $S$ is a pair, for it to give rise to a pole, it must also contain a constant line. In other words, it
must be a crystal. That is why crystals are the basic integration units. 

In order to reach the coincident zeros in $S$, the simplest way is to make a scaling change of variables, 
$\s_{x,x+1}=s\s_{x,x+1}'$, and let $s\to 0$. To get $m$ variables after the change, not $m\+1$, we must impose
a constraint on the $m$ quantities $\s'_{x,x+1}$. A convenient way to do so is to choose some $p\in S$ and set $\s'_{pr}=1$.
Line $p$ will be referred to as a {\it trigger}, the relation $\s'_{pr}=1$
will be called the {\it triggering relation}. 

The integration measure in $S$  in the new variables is
\be \prod_{x\in S\backslash r}d\s_x=s^{m-1}ds\prod_{x\in S\backslash r,p}d\s_x'.\labels{measure}\ee
$s$ will be taken as the integration variable for the crystal $S$. In this way, every crystal is responsible for one integration. 
Since there are $(n\-3)$ integrations in \eq{m}, at least that many crystals should be involved.
These crystals must either contain
no common lines, or else the lines of one crystal must be completely inside the other. 
This is so because the pole in each integration arises when all the $\s$-variables approach $\s_r$ of
its defect. Consider two crystals with defects $r$ and $t$ respectively. If $x$ is a line common to the two
crystals, then the two poles from the two crystals occur when $\s_x\to\s_r$ and $\s_x\to\s_t$. This is impossible
if $r\not= t$, so two crystals either do not overlap, or one must be inside another, with a common defect. 

If two crystals do not intersect, it does not matter which integration is carried out first. If one crystal is inside the other,
it is much more convenient to carry out the integration for the larger crystal before doing the one for the smaller crystal.

By power counting it is easy to see that there is no pole at $s=\infty$. 
This would not be the case if we had set a M\"obius constant to be infinity.
Whether a pole materializes
at  $s=0$ depends on the behavior of the integrand. As $s\to 0$,
the scattering functions become
\be f_x&\to& {1\over s}\sum_{y\in S\bsl x}{2k_x\.k_y\over\s'_{xy}}={1\over s}(f^S_x)',\quad (x\in S\backslash r)\nn\\
f_a&\to&\sum_{b\in T\bsl a}{2k_a\.k_b\over\s_{ab}}+{2k_a\.k_{\bar q}\over\s_{a\bar q}}=f^{\bar T}_a,\quad (a\in T\backslash s,t),\labels{fs}\ee
where $(f^S_x)'$ is  the partial scattering function defined in \eq{psf}, with $\s$ replaced by $\s'$, and $f_a^{\bar T}$
is the partial scattering function for the set $\bar T$ defined below.
The set $T$
is the complement of $S$ in $A$, namely $T=A\bsl S$, consisting of the remaining $n\-m\-1$ lines, which necessarily include the remaining two
\M constant lines $s$ and $t$. The letter $\bar q$ stands for an extra off-shell line with
momentum $k_{\bar q}=\sum_{x\in S}k_x=k^T$ (see \S~III for the definition of $k^T$). Moreover,
$\s_{\bar q}=\s_r$ because  $\s_x\to\s_r$ for all $x\in S$ as $s\to 0$. The set $\bar T=T\cup\{\bar q\}$ has this extra line $\bar q$
added to the set $T$. See Fig.~3.

Note that there is an asymmetry between $S$ and $T$, and this asymmetry will persist throughout. The asymmetry comes about
because $S$ is a crystal, but $T$ may not be.

From \eq{fs} we see that the factor $\prod_{i\not=r,s,t}f_i^{-1}$ in \eq{m} is proportional to $s^m$. Taking \eq{measure} into account, the integrand
has a simple pole at $s=0$ if and only if $\s_{(\a)}\s_{(\b)}$ goes to zero like $s^{2m}$, which is the case if and only if $S$ is
a crystal, as asserted previously.

If $r\in S$ were not a constant line, then $1/f_r$ would be present in the integrand of \eq{m}. Its scaling change under \eq{fs}
would bring along an extra factor $s$ to the integrand, thereby destroying the pole. 
This confirms that $r$ must be a constant line.

Let us see what becomes of the dynamical factor when $s\to 0$.
Recall that the colors $\a$ and $\b$ are cyclic, so they can be written in the form $\a=(\a_S,\a_T)$ and $\b=(\b_S,\b_T)$, which
can be used to define two other ordered sets $\a_T$ and $\b_T$. 
We shall refer to $\a_S, \b_S, \a_T, \b_T$ as {\it partial colors}.

For example, suppose $\a=(123456789)$ and $\b=(124395786)$. For the pairing with $S=\{1234\}$, we see from
Fig.~1 that  $T=\{56789\}$,
$\a_S=\bk{1234}$,  $\b_S=\bk{1243}$, $\a_T=\bk{56789}$
and $\b_T=\bk{95786}$.

In the limit $s\to 0$,   we can write (see \eq{sigma} for definitions)

\be
\s_{(\a)}&=&\s_{(\a_S\a_T)}=s^m\s'_{[\a_S]}\s_{(\bar q\a_T)}=s^m\s'_{[\a_S]}\s_{(\a_{\bar T})},\nn\\
\s_{(\b)}&=&\s_{(\b_S\b_T)}=s^m\s'_{[\b_S]}\s_{(\bar q\b_T)}=s^m\s'_{[\b_S]}\s_{(\b_{\bar T})}.\labels{ss}\ee
Though not immediately useful, the following relations are also true:
\be
\s_{[\a]}&=&\s_{[\a_S\a_T]}=s^m\s'_{[\a_S]}\s_{[\a_{\bar T}]},\nn\\
\s_{[\b]}&=&\s_{[\b_S\b_T]}=s^m\s'_{[\b_S]}\s_{[\b_{\bar T}]}.\labels{ss2}\ee

Collecting \eq{measure}, \eq{fs}, \eq{ss}, and remembering that $\s_{\bar q}=\s_r$, we can evaluate the $s$-integration in $\eq{m}$ to get

\be m(\a|\b)&=&J_S\.J_{\bar T},\labels{fac1}\\
J_S&=&\({-1\over 2\pi i}\)^{m-1}\oint_{\Gamma_S}\(\prod_{x\in S\backslash r,p}{d\s'_x\over (f^S_x)'}\){1\over (f^S_p)'}{1\over\s'_{[\a_S]}\s'_{[\b_S]}},\labels{js}\\
J_{\bar T}&=&\({-1\over 2\pi i}\)^{n-m-3}\oint_{\Gamma_T}\s^2_{(rst)}\(\prod_{a\in \bar T\backslash s,t,\bar q}{d\s_a\over f^{\bar T}_a}\){1\over\s_{(\a_{\bar T})}\s_{(\b_{\bar T})}},\labels{jt}\ee
where $\Gamma_S$ encircles  $(f^S_x)'=0$ for all $x\in S\bsl r,p$, and $\Gamma_T$ encircles $f^{\bar T}_a=0$ for all
$a\in \bar T\bsl s,t,\bar q$.  Using \eq{fpr}, $(f_p^S)'=(k^S)^2/\s'_{pr}=(k^S)^2$ because $\s'_{pr}=1$. 
If we add to $S$ an extra line $q$
to soak up all its missing momentum, then $k_q^2=(k^S)^2=(f^S_p)'$.
With that, and defining $\bar S=S\cup\{q\}$ to include the extra line $q$ in $S$, we can rewrite $J_S$ as 
\be
J_S=\({-1\over 2\pi i}\)^{m-1}\oint_{\Gamma_S}\(\prod_{x\in \bar S\backslash r,p,q}{d\s_x\over f^S_x}\){1\over\s_{[\a_S]}\s_{[\b_S]}}{1\over k_q^2}\equiv J_{\bar S}{1\over k_q^2},\labels{js2}\ee
where the integration variables $\s'$ have been renamed $\s$. In this form, the propagator is explicit and 
the factorization relation \eq{fac1} reads  (see Fig.~3)
\be m(\a|\b)=J_{\bar S}{1\over k_q^2}J_{\bar T},\labels{fac2}\ee
but please note the asymmetry between $J_{\bar S}$ and $J_{\bar T}$. We shall refer to $J_{\bar S}$ and $J_{\bar T}$ as 
{\it partial amplitudes}.

\section{An Example}
To illustrate the formalism of previous sections, the $n=5$ amplitude with diagonal colors
is computed here. This amplitude has been computed elsewhere by other means \cite{CHY3,DG1,Kal}.

We shall take the colors to be $\a=\b=(12345)$, and the constant lines to be $r,s,t=1,3,5$. Then \eq{m} becomes
\be M:=m(\a|\a)=\(-{1\over 2\pi i}\)^2\oint_\Gamma {\sigma_{(135)}^2d\sigma_2d\sigma_4
\over f_2f_4\sigma_{(12345)}\sigma_{(12345)}},\labels{m5}\ee
with
\be 
 f_2&=&{s_{21}\over \sigma_{21}}+{s_{23}\over \sigma_{23}}+{s_{24}\over \sigma_{24}}+{s_{25}\over \sigma_{25}}, \nn\\
f_4&=&{s_{41}\over \sigma_{41}}+{s_{42}\over \sigma_{42}}+{s_{43}\over \sigma_{43}}
    +{s_{45}\over \sigma_{45}},\nn\\
\s_{(135)}&=&\s_{13}\s_{35}\s_{51},\nn\\
\s_{(12345)}&=&\s_{12}\s_{23}\s_{34}\s_{45}\s_{51}.   \labels{m52} 
 \ee
 It will also be convenient to define, for any $n$,
 \be 
s_{ij\cdots k}&=&(k_i+k_j+\cdots +k_k)^2. \labels{sk}\ee

Before starting, let us review some of the salient features discussed above. Contributions to the two integrations in \eq{m5}
come from two crystals, either non-intersecting, or one completely inside the other. Equations \eq{measure} to
\eq{fac2} will be used to carry out the integrations. 

The first task is to identify these two compatible crystals from the
 five pairing diagrams shown in Fig.~2, and repeated in Fig.~4 with constants displayed as dashed lines.
There are five crystals, 
$S_a=\{12\},\  S_b=\{23\},\ S_c=\{234\},\ S_d=\{34\}$, and  $S_e=\{45\}$, each residing in two of the five diagrams. 
The compatible pairs are  $A=(S_a,S_e),\ B=(S_b,S_c),\ C=(S_a,S_d),\ D=(S_b,S_e),\ E=(S_c,S_d)$, corresponding
to the five diagrams shown in Fig.~3. The pairs in $A, C, D$ are non-intersecting, so it does not matter which
integration is first carried out. The pairs $B$ and $E$ have one crystal inside the other, so integrations of the larger
crystal $S_c$ should be carried out first. 

\bc \igw{9}{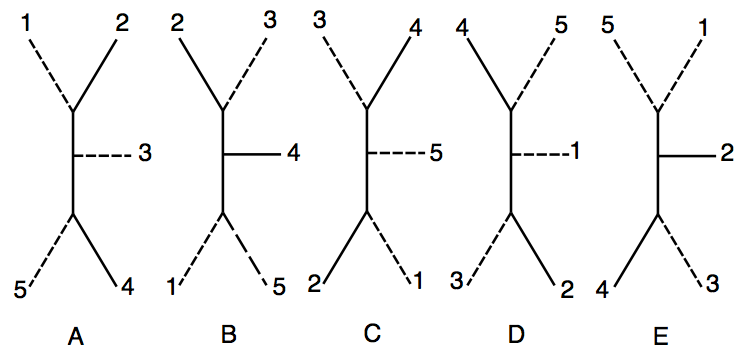}\\ Fig.~4.\quad The five pairing diagrams for $\a=\b=(12345)$. Dashed lines are constant lines\ec 

For the non-intersecting pairs, integration of each of the two non-intersecting crystals can be carried out independently.
$S_a$ yields the pole at $\s_{21}=0$, $S_b$ yields the pole at $\s_{23}=0$, $S_d$ yields the pole at $\s_{43}=0$, and 
$S_e$ yields the pole at $\s_{45}=0$. Evaluating the residues at these poles lead to
\be
M_A&=&-{1\over 2\pi i}\oint_{\Gamma_4} {\sigma_{(135)}^2d\sigma_4
\over s_{21}f_{4a}\sigma_{(1345)}\sigma_{(1345)}}={1\over s_{12}s_{45}},\nn\\
&&f_{4a}={s_{41}+s_{42}\over \sigma_{41}}+{s_{43}\over \sigma_{43}}+{s_{45}\over \sigma_{45}};\nn\\
M_C&=&-{1\over 2\pi i}\oint_{\Gamma_4} {\sigma_{(135)}^2d\sigma_4
\over s_{21}f_{4a}\sigma_{(1345)}\sigma_{(1345)}}={1\over s_{12}s_{34}};\nn\\
 M_D&=&-{1\over 2\pi i}\oint_{\Gamma_4} {\sigma_{(135)}^2d\sigma_4
\over s_{23}f_{4b}\sigma_{(1345)}\sigma_{(1345)}}={1\over s_{23}s_{45}},\nn\\
&&f_{4b}={s_{41}\over \sigma_{41}}+{s_{42}+s_{43}\over \sigma_{43}}+{s_{45}\over \sigma_{45}},\nn\\
\ee  

In the remaining two cases, we should first carry out the integration for the larger crystal $S_c$. The difference
between the two is that in one case, the second integration involves $S_b$ with a pole at $\s_{23}=0$, and in the other
case, the second integration involves $S_d$ with a pole at $\s_{43}=0$. In the first case, $S_b$, we should take 
$p=4, \s'_{43}=1$, and in the second case, $S_e$, we should take $p=2, \s'_{23}=1$.
In this way we get
\be
M_B&=&-{1\over 2\pi i}\oint_{\Gamma_2}{\sigma_{135}^2d\sigma'_2\over f_{2c}f_{4c}\sigma_{135}^2(\sigma'_{23}\sigma'_{34})^2}, \nn\\
&&f_{2c}={s_{23}\over\sigma'_{23}}+{s_{24}\over\sigma'_{24}},\quad 
 f_{4c}={s_{42}\over\sigma'_{42}}+{s_{43}\over\sigma'_{43}}, \ee    
 where $\Gamma_2$ encloses $f_{2c}=0$ counter-clockwise. 
With $\s'_{43}=1$, we
can use $f_{2c}=0$ to solve for $\s'_{42}$ to get $\s'_{42}=s_{24}/(s_{23}+s_{24})$. Substituting this into $f_{4c}$
gives $f_{4c}=s_{234}$. Now we can carry out the $\s'_2$-integration  at the pole defined by $f_{2c}=0$ to get 
\be M_B=-{1\over 2\pi i}\oint_{\Gamma_2}{d\sigma'_2\over f_{2c}s_{234}{\sigma'_{23}}^2}={1\over s_{51}s_{23}}.\ee

$M_E$ is obtained in exactly the same way to get 
\be
M_E={1\over s_{34}s_{51}}.\ee
The total five-point amplitude is therefore
\be M=M_A+M_B+M_C+M_D+M_E={1\over s_{12}s_{45}}+{1\over s_{23}s_{51}}+{1\over s_{34}s_{12}}+{1\over s_{45}s_{23}}+{1\over s_{51}s_{34}}.\labels{m5result}\ee

In this calculation, we have chosen to obtain the two propagators in each of the five diagrams by elementary integrations. 
We can also get them using \eq{fpr}, as is done in \eq{js2}.

\section{Factorization as Crystal Fracture}

Before proceeding further, let us summarize the mathematical operations of \S VII by in intuitive language,
using the previously defined concepts of `crystal', `defect', and `trigger'. This language would also be helpful
in visualizing subsequent integrations.

The integration resulting in  Fig.~3 and \eq{fac2} amounts to separating the crystal $S$ away from the rest of the
amplitude. We might think of this as a cleavage of the crystal $S$, triggered by pulling the line $p$, and that this
may happen because the crystal is already weaken by the presence of a defect $r$. As a result of the operation,
 line $p$ is pulled away to become
 the internal line (propagator) separating $S$ from the rest, thereby leaving behind an empty slot which is another defect in $S$.
 
 A defect $r$ was defined to be a constant line, but it can also be understood as one where $f_r$ is absent in the integrand.
 It is in this second sense that $p$ is another defect.

\section{Subsequent Integrations}
If the amplitude contains $(n\-3)$ mutually 
compatible crystals, namely, crystals that are non-intersecting or one  inside another, 
then all the integrations in the amplitude can be performed and the final result obtained. 
This is the case for the $n=5$ amplitude discussed in \S IX.
However, there are cases, 
the nine-point amplitude to be discussed in \S XI for example, where there are not enough compatible crystals.
Then how should the subsequent integrations be carried out?

There are not enough crystals because there are not enough defects. At the beginning there are only
three defects, $r,s,t$, but after a crystal $S$ is pried away from the rest, another defect $p$ is created. 
With a new defect, new crystals can  be created to provide further integrations. 

Nevertheless, there are complications associated with the subsequent integrations that requires consideration. To start with, although
the `partial amplitudes' $J_{\bar S}$ and $J_{\bar T}$ in \eq{js2} and \eq{jt} resemble the CHY amplitude \eq{m},
they are not the same. It is true that
in each case the scattering particles are contained in a set with zero total momentum, $\bar S$ for $J_{\bar S}$, $\bar T$
for $J_{\bar T}$, and $A$ for \eq{m}. Moreover,
each of these sets contains three constant lines, $r,p,q$ in 
 $J_{\bar S}$,  $s,t,\bar q$ in $J_{\bar T}$, and  $r,s,t$ in \eq{m}.
But beyond these similarities there are many differences.
Whereas each $f_i$ in \eq{m} is given by a sum over $j\in A\bsl i$, 
each  $f_x^S$ in $J_{\bar S}$ is given by a sum over $y\in S\bsl x$, not $y\in\bar S\bsl x$.
The term $2k_x\.k_q/\s_{xq}$ is missing in the sum, though a similar term $2k_a\.k_{\bar q}/\s_{a\bar q}$ is present in $f_a^{\bar T}$
of $J_{\bar T}$. See \eq{fs}. However, there is a difference
between $f_a^{\bar T}$ in $J_{\bar T}$ and $f_i$ in \eq{m}. $k_{\bar q}$ is off-shell, so $2k_a\.k_{\bar q}\not=(k_a+k_{\bar q})^2$, but 
every line in \eq{m} is on-shell with $2k_i\.k_j=(k_i+k_j)^2$.

 Dynamical factors are also different. The analog of the dynamical factors in \eq{m}
would be $\s_{(\a_{\bar S})}\s_{(\b_{\bar S})}$ and $\s_{(\a_{\bar T})}\s_{(\b_{\bar T})}$, but instead, it is
  $\s_{[\a_{ S}]}\s_{[\b_{S}]}$ in $J_{\bar S}$, though still $\s_{(\a_{\bar T})}\s_{(\b_{\bar T})}$ in $J_{\bar T}$.
Lastly, the analog of the normalization factor $\s_{(rst)}^2$ is present in $J_{\bar T}$ but absent in $J_{\bar S}$.

These points can be summarized by saying that  partial amplitudes and  CHY
amplitudes are essentially the same in  aspects when the off-shell lines $q,\bar q$ are not involved, 
otherwise they are different and we must be careful. We will show that partial amplitudes can be
integrated in much the same way as the original amplitude because their off-shell lines $q,\bar q$
are never involved.

Integration relies on crystals, which are defined via pairs.
Pairs were originally defined by matching the two colors, $\a$ and $\b$, but they can also be defined
by matching $\s_{(\a)}$ and $\s_{(\b)}$. In fact, it is this latter matching that is relevant to integration,
because they determine the degree of zeros of $\s_{(\a)}\s_{(\b)}$, and hence the location of the $s$-poles.
In $J_{\bar S}$, pairs are determined by the matching of $\s_{[\a_S]}$ and $\s_{[\b_S]}$. As such,
the off-shell line $q$ which enters only into $\s_{(\a_{\bar S})}$ and $\s_{(\b_{\bar S})}$ is not involved, as claimed.
In $J_{\bar T}$, the dynamical factors are $\s_{(\a_{\bar T})}$ and $\s_{(\b_{\bar T})}$.
Although the off-shell line $\bar q$ is present, it is inactive as far as pairing is concerned. 
 This is so because $\s_{\bar q}=\s_r$, so if $\bar q$ is involved in forming a pair in $\bar T$, it must enter as the defect
 of a crystal. 
This is not allowed because $r$ is already a defect in $S$, and two crystals are not allowed to have the same defect
unless one is inside the other. 

In short, pairs in $J_{\bar S}$ and $J_{\bar T}$ are simply the pairs in the original CHY amplitude that reside
completely in $\bar S$ or in $\bar T$. Since a new defect $p$ is present in $\bar S$, there may be new crystals
in $\bar S$ not present originally. In terms of pairing diagrams, pairs in the original amplitude are formed from the external lines 
to one side of an internal line. If an internal line is cut to separate it into a crystal $S$ and the remaining part $T$, then
the pairs in each can still be formed by the external lines to one side of an internal line, but that side must be the side that
does not involve the off-shell line. For example, take the pairing diagram in Fig.~1. If $S=\{91234\}$ and $T=\{5678\}$, 
then the internal line between line 9 and line 5 is $q$ on the $S$ side and $\bar q$ on the $T$ side. On the $S$ side, 
$\{1234\}$ forms a pair in $\bar S$, but not $\{q9\}$, because the latter involves the off-shell line $q$.

Having thus identified the crystals of the partial amplitudes, the next integration would
base on one of these crystals, be it in $\bar S$ or in $\bar T$. We must pick a new trigger $p'$ in that crystal, distort
the contour away from $\tilde f_{p'}=0$ to surround the dynamical factors. $\tilde f$ is the scattering function appropriate
to the partial amplitude, which is not quite the same as the original
scattering function. In order to be able to use \eq{fpr} to turn it into an inverse propagator to complete the factorization
of the partial amplitude, we must show that we can replace $\tilde f_{p'}$ by $f^S_{p'}$ or $f^T_{p'}$, whichever is appropriate.

We know from \eq{js} that in $\bar S$, $\tilde f_{p'}$ is indeed $f^S_{p'}$, but from \eq{jt}, we see that in $\bar T$,
$\tilde f_{p'}=f^{\bar T}_{p'}$, not $f^T_{p'}$. However, the additional off-shell term involves $\bar q$,
and as argued before, $\bar q$ can never be in a crystal in $\bar T$, so this term in the scattering function is irrelevant when $s\to 0$. In
other words, the subsequent integration based on another crystal can be carried out 
in the same way as before, to factorize the partial amplitude
and to expose another propagator. This procedure can be repeated until all the propagators of the Feynman diagrams
are exposed.

One last remark. If we define a constant line $c$ to be one that $f_c$ is not present in the integrand, then
both defects and off-shell lines are constants, but  defect lines are on-shell.

\section{A nine-point amplitude}
To illustrate how the whole thing works, let us turn to the example shown in the pairing diagram Fig.~1,
with $\a=(123456789)$ and $\b=(124395786)$. The three constant lines $r, s, t$ are shown as
dashed lines in the three different cases of Fig.~5. They are the initial crystal defects. 

\bc\igw{10}{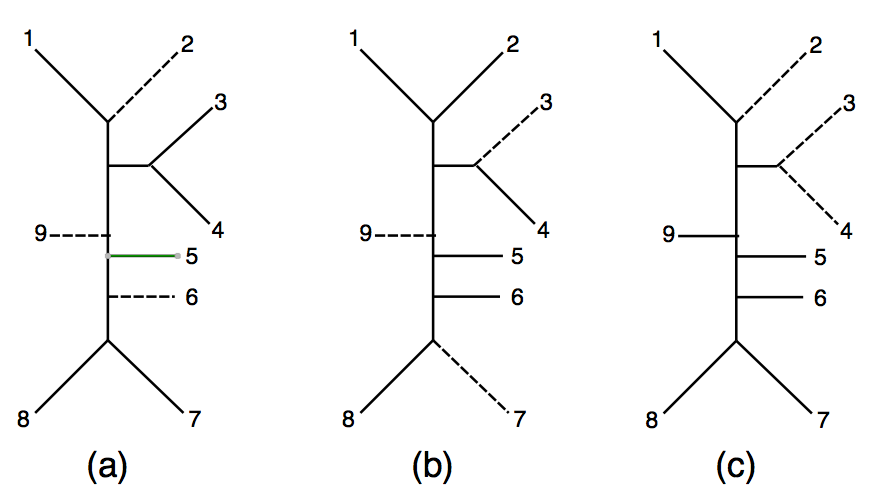}\\ Fig.~5\quad Three different choices of constant lines (dashed) for Figs.~1(a) and 1(b)\ec

 The formation of crystals depend on the location of defects. With different defects, these three cases
contain different crystals $S$ as shown below:
\bn
\i Fig.~5(a).
	\bn
	\i If 2 is the defect, $S$ could be $\{12\}$ or $\{1234\}$.
	\i If 6 is the defect, $S$ could be $\{678\}$ or $\{5678\}$.
	\i There is no crystal that contains 9 as its only defect, so factorization cannot be carried out
	with defect 9.\en
\i Fig.~5(b).
	\bn	
	\i If 3 is the defect, $S$ could be $\{34\}$ or $\{1234\}$.
	\i If 7 is the defect, $S$ could be $\{78\}, \{678\}$, or $\{5678\}$.
	\i There is no crystal that has 9 as its only defect.\en
\i Fig.~5(c).
	\bn
	\i If 2 is the defect, $S$ has to  be $\{12\}$.
	\i There is no crystal containing either 3 or 4 as the only defect.
	\i Consequently this is a bad choice of $r,s,t$. After factorizing out the crystal $\{12\}$,
	it is impossible to carry out further factorizations using the method discussed above.\en
\en

We will now concentrate on the case of Fig.~5(a), shown again in Fig.~6, but
with the six propagators labelled A to F.

\bc\igw{3}{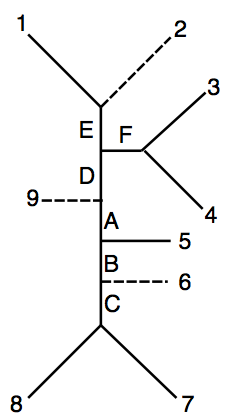}\\Fig.~6\quad Same as Fig.~5(a) but with names assigned to the propagators\ec

Given a defect, any variable line inside the crystal can be used to trigger the cleavage. 
The trigger turns into a new defect in the fractured crystal,
so there are many 
possible combinations
of successive choice of defect and trigger lines. In what follows we just choose one random
combination to illustrate the process. 
A different combination would lead to a different way of fracturing the whole crystal into the final crystals.

To simplify writing and make it easier to read, we will abbreviate $\s_{[\cdots]}$ as $[\cdots]$ and $\s_{(\cdots)}$ as  $(\cdots)$.
An arrow $\longrightarrow$ signifies factorization after an integration, and the notation $pr$ above the arrow reminds us 
to impose the trigger relation $\s_{pr}=1$. The first letter $p$ is the trigger, the second letter $r$ is the defect.
A dot on top of a line number indicates that line to be a constant line. If the constant line is on-shell, then
it is a defect.

Here is the computation:
\be
(1\sb 2345\sb 678\sb 9)(1\sb 243\sb 9578\sb 6)&\sa{56}&[\ \sb 5\ \sb 678][\ \sb 578\sb 6\ ]\sp{A}(\ \sb 91\2 34\sb 6\ )
(1\2 43\sb 9\ \sb 6\ )\labels{e1}\\
&\sa{76}&[\ \sb 6\ \sb 78][\ \sb 78\sb 6\ ]\sp{B}[\ \sb 5\ \sb 6\ ][\ \sb 5\ \sb 6\ ]\sp{A}(\ \sb 91\2 34\sb 6\ )
(1\2 43\sb 9\ \sb 6\ )\labels{e2}\\
&\sa{87}&[\ \sb 7\ \sb 8\ ][\ \sb 7\ \sb 8\ ]\sp{C}[\ \sb 6\ \sb 7\ ][\ \sb 7\ \sb 6\ ]\sp{B}\sp{A}(\ \sb 91\2 34\sb 6\ )
(1\2 43\sb 9\ \sb 6\ )\labels{e3}\\
&\sa{42}&-\sp{C} \sp{B} \sp{A}[1\2 3\4\ ][1\2\ \4 3]\sp{D}(\ \9\ \2\ \6\ )(\ \2 \  \9\ \6\ )\labels{e4}\\
&\sa{12}&-\sp{C} \sp{B} \sp{A}[\ \1\ \2\ ][\ \1\ \2\ ]\sp{E}[\ \2 3\4\ ][\ \2 \  \4 3\ ]\sp{D}(\ \9\ \2\ \6\ )(\ \2 \  \9\ \6\ )\labels{e5}\\
&\sa{34}&-\sp{C} \sp{B} \sp{A}\sp{E}[\ \3\ \4\ ][\ \4\ \3\ ]\sp{F}[\ \2 \ \4\ ][\ \2 \  \4 \ ]\sp{D}(\ \9\ \2\ \6\ )(\ \2 \  \9\ \6\ )\labels{e6}\\
&=&\sp{C} \sp{B} \sp{A}\sp{E}\sp{F}\sp{D}(\ \9\ \2\ \6\ )(\ \2 \  \9\ \6\ )\nn\\
&=&-\sp{C} \sp{B} \sp{A}\sp{E}\sp{F}\sp{D}(\ \9\ \2\ \6\ )(\ \9 \  \2\ \6\ ).\labels{e7}
\ee
The computational details are summarized in Fig.~7, in the $\a$-perspective on the left and the $\b$-perspective on
the right. The diagrams are a bit crowded, but after understanding what the symbols mean, they do summarize
all the computational details.
\bc\igw{7}{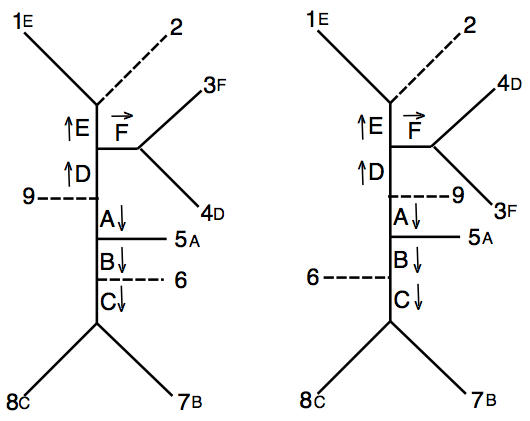}\\ Fig.~7.\quad Same as Fig.~6 but with algebraic details incorporated\ec

Here is how Fig.~7 should be read.
The order of integrations is $A,B,C,D,E,F$. The arrows at each propagator indicates which side is $S$, namely, the
crystal that is to be pried off. The letter besides a variable line tells the integration at which this line is taken to be the trigger $p$. 
For each $p$, the defect $r$ such that $\s_{pr}=1$ can also be read off from the diagram by remembering that $r$ may be
a dotted line in $S$, or a previous trigger line. For example, in the $B$ integration, $p=7$. On the arrow side of the propagator
$B$, the only possible defect is 6, so $r=6$. In the $C$ integration, $p=8$. There is no dotted line on the arrow side of 
propagator $C$, but there is a previous trigger line  7, so $r=7$.

Having determined the triggering relations, we can now proceed to determine the final $\s$-factor in the following way. 
Let us do so by following the integration order. Start with $A$, whose trigger is 5, and defect is 6. Remembering that
$\s_{(\a)}$ is given by the external lines on the left diagram, read counter-clockwise, and $\s_{(\b)}$ is given by the external lines
of the right diagram, also read counter-clockwise. On the left, $\s_{(\a)}$ contains $\s_{56}$, which is 1 by the triggering relation.
On the right, 5 and 6 are not consecutive, but subsequent integrations $B$ and $C$ would dissolve the lines 7 and 8 in between
because their $\s$'s would become 0 as $s\to 0$, hence eventually we will also have the factor $\s_{56}=1$. Next, look at integration
$B$ whose trigger is 7 and defect is 6. On the left, 6 and 7 are consecutive so we get $\s_{67}=-1$ immediately. On the right,
after dissolving line 8 by a subsequent integration $C$, we get $\s_{67}=1$, so the product of the two is $-1$.  Proceeding
thus, we see that each integration eliminates the $\s$ of the trigger and replace it by $\pm 1$, with the sign determined by
whether the (trigger, defect) pair are in the same order in both diagrams ($+1$), or opposite order ($-1$). At the end of the day,
the $\s$'s of variable lines all disappear, leaving behind from $\s_{(\a)}\s_{(\b)}$ a factor $\pm\s_{(rst)}^2$, with the sign determined
by whether the three constant lines are in the same cyclic order or opposite cyclic order between the left diagram and the right.
This $\s$ factor cancels the normalization factor $\s_{(rst)}^2$ in the numerator of \eq{m}, making the final result independent of
the choice of the \M constants. The overall sign at the end is determined by how many odd permutations there are to get the lines
on the left to be in the same order as the lines on the right, so it is the product of the signatures of $\a\in S_n$ and $\b\in S_n$.

 The diagram on the right of Fig.~7 is needed just to determine the $\s$ factors and the overall sign. Once we are convinced
that all the $\s$ factors eventually get cancelled out, leaving behind a sign which can be determined by the relative signature
of the two colors $\a$ and $\b$, there is no need any more for the diagram on the right. 

As mentioned before, there are many ways to choose the order of integrations and the trigger lines for each integration.
Fig.~8 gives another example to do it differently. However, it should be noted that the triggers must be chosen to ensure a complete set
of compatible crystals are available, so it is not totally random. For example, in Fig.~8, if the trigger line for
the $A$ integration were chosen to be line 1, then there is no way to extract the propagator $C$.

\bc\igw{3}{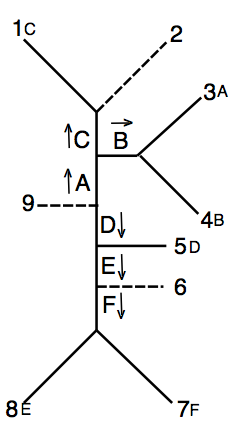}\\ Fig.~8.\quad Another way to do the integrations for the nine-point amplitude\ec

\section{Partial amplitude, off-shell amplitude, and shifted amplitude}
It was pointed out in \S~IX that the partial amplitudes $J_{\bar T}$ and $J_{\bar S}$ in \eq{jt} and \eq{js2} 
are different from the CHY on-shell amplitude \eq{m}. They are also different from the M\"obius-invariant CHY off-shell amplitude
\cite{LY1}.
Nevertheless, it will be shown in this subsection that they could effectively be,
though not exactly,  
the same as the shifted amplitudes used in BCFW factorization. This agrees with the result of Ref.~\cite{DG1}.

Consider the CHY amplitude \eq{m} for the particles in set $\bar S$, 
with $r,p,q$ as constant lines, but with the momenta of  $r$ and $q$ replaced by the shifted momenta
$\hat k_r=k_r+z\ell$ and $\hat k_q=k_q-z\ell$. The vector $\ell$ is a complex light-like
momentum orthogonal to $k_r$ (and $k_s$ of set $T$), so that $\hat k_r^2=0$ (and also $\hat k_s^2=0$). 
The complex number $z$ is chosen so that $\hat k_q^2=0$, namely,
$z=k_q^2/2k_q\.\ell$. With this choice, all the momenta in $\bar S$ are on-shell, and momentum conservation is preserved.
The CHY amplitude for $\bar S$ with these shifted momenta will be denoted by $\hat I_{\bar S}$,
and its scattering functions be denoted as $\hat f_x$.

Since $\hat I_{\bar S}$ is \M invariant, we may choose the \M constants $\s_r, \s_p, \s_q$ to be anything we like. 
Suppose we choose $\s_{pr}=1$ and $\s_q=\infty$. The normalization and dynamical factors in $\hat I_{\bar S}$ are $\s_{rpq}^2/\s_{(\a_{\bar S})}\s_{(\b_{\bar S})}$,
but with $\s_{pr}=1$ and $\s_q=\infty$, this becomes $1/{\s_{[\a_{S}]}}\s_{[\b_{ S}]}$, which is what appears 
in $J_{\bar S}$ in \eq{js2}.
The scattering functions $\hat f_x$ for $\hat I_{\bar S}$ and the scattering functions $f^S_x$ of $I_{\bar S}$
are almost identical too, but not quite. 
The former contains a term $2k_x\.\hat k_r/\s_{xr}$,
but the corresponding term in $f_x^S$ is $2k_x\.k_r/\s_{xr}$.
Such a difference is irrelevant if  further scaling and factorization of $S$ does not involve line $r$.
This would be the case if we choose the next defect 
to be $p$ and not $r$,   then $r$ cannot be involved because only one defect is allowed in each fracture, and the
difference in a term involving $r$ becomes irrelevant. 
For this to succeed it is crucial that
the crystal $S$ does contain two defects, $r$ and $p$, and that $p$ is not shielded by $r$ so that we can 
choose a smaller crystal containing the defect $p$ but not the defect $r$.
With that choice, $\hat I_{\hat S}$ and $J_{\bar S}$ are effectively the same at the end. 

Similar but not identical arguments also apply to $\hat I_{\bar T}$, defined to be the CHY amplitude for particles in $\hat T$,
with constant lines $s,t,\bar q$ and shifted momenta $\hat k_s=k_s-z\ell$ and $\hat k_{\bar q}=k_{\bar q}+z\ell=-\hat k_q$.
By choosing $\s_{st}=1$ and $\s_{\bar q}=\infty$, the normalization and dynamical factors of $\hat I_{\bar T}$ become
identical to those of $J_{\bar T}$, and the scattering functions $\hat f_a$ of $\hat I_{\bar T}$ are almost identical
with the scattering functions $f_a^{\bar T}$ of $J_{\bar T}$. With $\s_{\bar q}=\infty$, $f_a^{\bar T}=f_a^T$, so
the only difference between $\hat f_a$ and $f_a^{\bar T}$ is in the terms $2k_a\.\hat k_s/\s_{as}$ and $2k_a\.k_s/\s_{as}$.
Again this difference is irrelevant if subsequent fractures of crystal $T$ chooses $t$ as the defect rather than $s$.

Although it is equivalent to express the factorization in partial amplitudes and in the BCFW on-shell amplitudes, for further
factorization there is a distinct advantage to use the partial amplitudes. This is because the BCFW amplitudes are expressed
in terms of the shifted momenta, whereas the partial amplitudes are expressed in the original momenta. The propagators that
emerge from further factorization are given by the inverse sum of the original momenta, not the shifted momenta, so if
the BCFW amplitudes are used, there is a further task of showing that the shifted momenta also give rise to the right propagators. 

The shifted momentum on one side of the BCFW factorization is related to a momentum on the other side, in that sense
the factorization is somewhat non-local. This is needed to make all its momenta on-shell, though as a result some momenta
become complex. 
The CHY formula also appears to be non-local before the $\s$-integrations,
but after they simply turn into Feynman diagrams, which are local with off-shell propagators and real momenta.

\section{Summary}
The purpose of this paper is to 
examine the evaluation of CHY double-color scalar amplitudes, one integration
at a time, to expose the precise role of the \M constants and the  scattering functions.
To start with, we need to produce a pairing diagram
such as Fig.~1 from the two colors $\a$ and $\b$.
These pairing diagrams resemble
Feynman diagrams, and indeed at the end of the day they turn out to be exactly the Feynman diagrams, but at this stage
they are simply a graphical way to represent correlations of the two colors. From the pairing diagram we
identify all possible `crystals', which are the set of external lines to one side of an internal line
provided it also contains one and only one on-shell constant line (a `defect'). 
Each crystal corresponds to an integration, and compatible integrations must come from compatible crystals, which are
non-intersecting crystals or one crystal inside another. 

Even if the pairing diagram does  not initially provide enough compatible 
crystals  to perform all the integrations, new crystals can be formed along the way to do the job
because each integration triggers the creation of a new defect. 
Each integration exposes one propagator of the Feynman diagram which comes from the inverse of a scattering
function, the `trigger'. All the dependence on the \M constants get cancelled out at the end, leaving behind an overall sign
determined by the relative signature of the two colors. 

Two detailed examples are provided
to illustrate the terminology and the method, one a five-point amplitude with diagonal colors, and the other a
nine-point amplitude with non-diagonal colors.

\appendix
\section{An Infinite \M Constant}
In this Appendix we use the $n=5$ amplitude to illustrate the asymmetry caused by letting one of the \M constants
to become infinite, and the fictitious momentum poles thus produced in the intermediate step. 

Suppose we let $\sigma_1=0,\  \sigma_3=\infty, \ \sigma_5=1$. Then the 5-point amplitude in \eq{m5} becomes
\be M=-\({-1\over 2\pi i}\)^2\oint {d\sigma_2 d\sigma_4\over f_2f_4\sigma_{12}^2
\sigma_{45}^2}, \labels{ (A-9)}\ee
with 
\be f_2={s_{21}\over \sigma_{21}}+{s_{24}\over \sigma_{24}}+{s_{25}\over \sigma_{25}},  \ \ \ \  f_4={s_{41}\over \sigma_{41}}+{s_{42}\over \sigma_{42}}+{s_{45}\over \sigma_{45}}. \ee
Note that an asymmetry has been created in that $s_{23}$ and $s_{43}$ are now absent from the integrand. They
can be obtained only by momentum conservation from the remaining $s_{ij}$'s.

First carry out the $\s_2$ integration by distorting the contour away from $f_2=0$, while keeping it still around $f_4=0$.
Similar to the discussion in \S VIII, there are three regions (`crystals') that contribute to the integral. They involve
 the set of lines
$S_a=\{12\},\  S_b=\{23\},\ S_c=\{234\}$, with the $\s_i$'s within each set approach one another.  Since $\s_3=\infty$, these regions are
(a) $\sigma _{21}\to 0 $,
 (b) $\sigma_2 \to \infty ,  \ \sigma_4$ is finite, and (c) $\sigma_2\to \infty $ and
 simultaneously  $\sigma_4 \to \infty.$ They give rise to
 \be M(a)={1\over s_{12}}{1\over 2\pi i} \oint {d\sigma_4\over f_4 \sigma_{45}^2}.\ee
 The $\sigma_4$ integration is obtained by writing 
 \be\sigma_{41}= \sigma_4, \ \sigma_{42}=\sigma_{41}+\sigma_{12 }\approx
 \sigma_4, \ \sigma_{45}=\sigma_4-1. \ee  
 Since $\sigma_{12}\to 0$ in the $\sigma_2$ integration, and therefore  
 
 \be f_4\approx {(s_{41}+s_{42}+s_{45})\sigma_4-(s_{41}+s_{42})\over \sigma_4(\sigma_4-1)}.  \ee
 Evaluating at the zero of $f_4$ gives $\sigma_4^0={s_{41}+s_{42}\over s_{41}
 +s_{42}+s_{45}}$ and 
 
 \be M(a)={s_{41}+s_{42}\over s_{12}s_{45}(s_{41}+s_{42}+s_{45})}
 ={1\over s_{12}s_{45}}+{1\over s_{12}s_{34}},\ee
 where we have used $s_{41}+s_{42}+s_{45}=-s_{34}$.

We proceed now to the computation of $M(b)$.  An extra minus sign is present because of the reversal of
contour to account for $\sigma_2 \to \infty$. In this region (b),
\be f_2\approx {s_{21}+s_{24}+s_{25}\over \sigma_2}\approx {-s_{23}\over \sigma_2}, \ \ f_4\approx 
{s_{41}\over \sigma_{41}}+{s_{45}\over \sigma_{45}}.  \ee 
The integration over $\sigma_2$ leads to

\be M(b)={-1\over s_{23}}{1\over 2\pi i}\oint {d\sigma_4(\sigma_4)\over 
((s_{41}+s_{45})\sigma_4-s_{41}) (\sigma_4-1)}, \ee
where we have written $\sigma_{41}=\sigma_4, \ \sigma_{45}=\sigma_4-1.$ 
The zero of $f_4$ is at $\sigma_4^0={s_{41}\over s_{41}+s_{45}}$ and its 
residue gives 
\be M(b)={s_{41}\over s_{23}(s_{41}+s_{45})s_{45}}={1\over s_{23}s_{45}}
-{1\over s_{23}(s_{41}+s_{45})}.   \labels{mbapp}\ee
Note that a fictitious pole in the momentum space appears at $s_{41}+s_{45}=0$.

Finally we come to region (c), where both $\s_2$ and $\s_4$ are large. Let us first enforce the 
requirement $f_4=0$.  Here 

\be f_4\approx {s_{41}+s_{45}\over \sigma_4}+{s_{42}\over \sigma_4-\sigma_2}
={(s_{41}+s_{42}+s_{45})\sigma_4-(s_{41}+s_{45})\sigma_2\over 
\sigma_4(\sigma_4-\sigma_2)}. \ee
Therefore, we must constrain the $\sigma_4$ in $f_2$ to

\be\sigma_4^0={s_{41}+s_{45}\over s_{41}+s_{42}+s_{45}}\sigma_2. \ee
Then

\be f_2\approx {s_{21}+s_{25}\over \sigma_2}+{s_{24}\over \sigma _2-\sigma_4}
={s_{21}+s_{25}-s_{23}\over \sigma_2}.  \ee
Thus, 

\be M(c)=\({1\over 2\pi i}\)^2 \oint {d\sigma_2\over 
(s_{21}+s_{25}-s_{43})\sigma_2}\oint {d\sigma_4\over (s_{41} +s_{45}+s_{42})\sigma_4
-(s_{41}+s_{45})\sigma_2}\({\sigma_4^0-\sigma_2\over \sigma_4^0}\).\nn \ee 
Carrying out the integrations, we arrive at

\be M(c)=-{s_{42}\over (s_{41}+s_{45})s_{43} s_{15}}, \ee
where we have used $s_{21}+s_{25}-s_{43}=-s_{15}, \ s_{41}+s_{42}+s_{45}=-s_{43}$ and ${\sigma_4^0-\sigma_2\over \sigma_4^0}=-{s_{42}\over s_{41}+s_{45}}$. A fictitious momentum pole also arises in this integral.

We now combine this result with the last term of $M(b)$ in \eq{mbapp} and 
note that 

\be Y\equiv-{1\over s_{41}+s_{45}}\({s_{42}\over s_{43}s_{15}}+{1\over s_{23}}\)
=-{1\over (s_{41}+s_{45})s_{23}s_{43}s_{15}}X,  \ee
where

\be X&=&s_{23}s_{42}+s_{43}s_{15}=s_{23}s_{42}+s_{43}(s_{23}+s_{24}+s_{34})\nn\\
&=&(s_{23}+s_{43})(s_{42}+s_{43})=- (s_{23}+s_{43})(s_{14}+s_{45}).\ee
Putting this into $Y$ results in

\be Y={s_{23}+s_{34}\over s_{15}s_{23}s_{34}}={1\over s_{15}s_{23}}+
{1\over s_{15}s_{34}}.  \ee
Altogether then, the sum of all three regions is the same as what is in \eq{m5result}. The fictitious pole
encountered in region (b) and region (c) disappear, leaving behind only the correct Feynman diagram poles.


\begin{thebibliography}{9}
\bibitem{CHY1} F. Cachazo, S. He, and E.Y. Yuan, Phys.~Rev.~D {\bf 90} (2014) 065001, arXiv: 1306.6575; 
\bibitem{CHY2} F. Cachazo, S. He, and E.Y. Yuan,	Phys.~Rev.~Lett. {\bf 113} (2014) 17161, arXiv: 1307.2199; 
\bibitem{CHY3} F. Cachazo, S. He, and E.Y. Yuan,	JHEP {\bf 1407} (2014) 033, arXiv: 1309.0885;
\bibitem{CHY4} F. Cachazo, S. He, and E.Y. Yuan,	JHEP {\bf 1501} (2015)121, arXiv: 1409.8256;
\bibitem{CHY5} F. Cachazo, S. He, and E.Y. Yuan,	JHEP {\bf 1507} (2015) 149, arXiv:1412.3479;		
\bibitem{CHY6} F. Cachazo, S. He, and E.Y. Yuan,	Phys.~Rev. D {\bf 92} (2015)  065030, arXiv:1503.04816.
\bibitem{ST1}    L. Mason and D. Skinner,  JHEP {\bf 1407} (2014) 048, arXiv:1311.2564
\bibitem{ST2}	N. E. J. Bjerrum-Bohr, P. H. Damgaard, P. Tourkine, and P. Vanhove, Phys.~Rev. {\bf D90} (2014) 106002, 
		arXiv: 1403.4553
\bibitem{ST3}	C. Baadsgaard, N. E.J. Bjerrum-Bohr, J.L. Bourjaily, P.H. Damgaard,		
		 JHEP {\bf 1509} (2015) 136, arXiv:1507.00997
\bibitem{LOOP1} 	Y. Geyer, L. Mason, R. Monteiro, P. Tourkine, Phys.~Rev.~Lett. {\bf 115} (2015) 121603, arXiv:1507.00321
\bibitem{LOOP2} 	S. He, E.Y. Yuan, arXiv:1508.06027
\bibitem{LOOP3}      C. Baadsgaard, N.E.J. Bjerrum-Bohr, J.L. Bourjaily, P.H. Damgaard, B. Feng, JHEP 1511 (2015) 080, arXiv:1508.03627
\bibitem{LOOP4}      B. Feng, arXiv:1601.05864
    
\bibitem{LY1}   C.S. Lam and Y-P. Yao, arXiv:1511.05050	
\bibitem{MASS1} S. G. Naculich,  JHEP {\bf 1409} (2014) 029, arXiv:1407.7836
\bibitem{MASS2}	 S. G. Naculich,	  JHEP{\bf 1505} (2015) 050, arXiv: 1501.03500 
\bibitem{MASS3}	 S. G. Naculich,	  JHEP {\bf 1509} (2015)122, arXiv:1506.06134
\bibitem{MASS4}	 S. Weinzierl,  JHEP{\bf 1503} (2015) 141, arXiv:1412.5993	
\bibitem{DG2}  L. Dolan, P. Goddard, JHEP {\bf 1407} (2014) 029, arXiv:1402.7374.	
\bibitem{CK15}  C. Cardona, C. Kalousios, arXiv:1511.05915.
\bibitem{DG3} L. Dolan, P. Goddard, arXiv:1511.09441.
\bibitem{Kal} C. Kalousios,JHEP {\bf 1505} (2015) 054, arXiv: 1502.07711 [hep-th].
\bibitem{FG} F. Cachazo, H. Gomez,  arXiv:1505.03571.
\bibitem{BBBD2} C. Baadsgaard, N. E. J. Bjerrum-Bohr, J. L. Bourjaily, and P. H. Damgaard,  
		JHEP {\bf 1509} (2015)136, arXiv:1507.00997.
\bibitem{DG1} L. Dolan and P. Goddard, JHEP {\bf 1401}(2014) 152, arXiv:1311.5200\\
\bibitem{BBBD} C. Baadsgaard, N. E. J. Bjerrum-Bohr, J. L. Bourjaily, and P. H. Damgaard, JHEP {\bf 1509} (2015)129, arXiv:1506.06137.
\bibitem{BCFW} R. Britto, F. Cachazo, B. Feng, E. Witten, Phys.~Rev.~Lett. {\bf 94} (2005) 181602, arXiv: hep-th/0501052.
\bibitem{KT} G. K\"all\'en and J. Toll, Helv.~Phys.~Acta 33 (1960) 753.
\bibitem{LYgauge} C.S. Lam and Y.P. Yao, to be published.
\bibitem{SE1} R. Monteiro and D. O'Connell,  JHEP {\bf 1403} (2014) 110, arXiv:1311.1151
\bibitem{SE2}	C. Kalousios, J.~Phys.~{\bf A47} (2014) 215402, arXfiv: 1312.7743	
\bibitem{SE3}	S. Weinzierl, JHEP {\bf 1404} (2014) 092, arXiv: 1402.2516
\bibitem{SE4}	L. Dolan and P. Goddard, JHEP {\bf 1407} (2014) 029, arXiv: 1402.7374
\bibitem{SE5}	Y-H. He, C. Matti, C. Sun, 	JHEP {\bf 1410} (2014) 135, arXiv:1403.6833	
\bibitem{SE6}	C. S. Lam, Phys.~Rev. {\bf D91} (2015) 045019, arXiv:1410.8184
\bibitem{SE7}	R. Huang, J. Rao, B. Feng, Y-H. He, 	arXiv:1509.04483
\bibitem{SE8}	M. Sogaard, Y. Zhang, arXiv:1509.08897
\bibitem{SE9}   C. Cardona, C. Kalousios, arXiv:1509.08908

\end{thebibliography}
\end{document}